# Synchronized Eruptions on Io: Evidence of Interconnected Subsurface Magma Reservoirs

A. Mura[1*], R. Lopes[2], F. Nimmo[3], S. Bolton[4], A. Ermakov[5], J. T. Keane[2], F. Tosi[1], F. Zambon[1], R. Sordini[1], J. Radebaugh[6], J. Rathbun[7], W. McKinnon[8], S. Goossens[9], M. Paris[1], M. Mirino[1], G. Piccioni[1], A. Cicchetti[1], R. Noschese[1], C. Plainaki[10], G. Sindoni[10]

[1] Institute for Space Astrophysics and Planetology, National Institute for Astrophysics (IAPS/INAF), Rome, Italy

[2] Jet Propulsion Laboratory, California Institute of Technology, Pasadena, CA, USA

[3] Department of Earth and Planetary Sciences, University of California, Santa Cruz, Santa Cruz, CA, USA

[4] Southwest Research Institute, San Antonio, TX, USA

[5] Department of Aeronautics and Astronautics, Stanford University, Stanford, CA, USA

[6] Brigham Young University, Provo, UT, USA

[7] Cornell University, Ithaca, NY, USA

[8] Department of Earth, Environmental, and Planetary Sciences, Washington University in St. Louis, St. Louis, MO, USA

[9] NASA Goddard Space Flight Center, Greenbelt, MD, USA

[10] Agenzia Spaziale Italiana, Rome, Italy

**Abstract**. On December 27, 2024, Juno's JIRAM infrared experiment observed an unprecedented volcanic event on Io's southern hemisphere, covering a vast region of ~ 65,000 km², near 73°S, 140°E. The total power output is estimated between 140 and 260 TW, potentially the most intense ever recorded, surpassing the brightest eruption at Surt in 2001 (~80 TW). Within that region, only one hot spot was previously known (Pfd454). This feature was earlier estimated to cover an area of 300 km² with a total power output of 34 GW. JIRAM results show that the region produces a power output of 140-260 TW, over 1,000 times higher than earlier estimates. Three adjacent hot spots also exhibited dramatic power increases: P139, PV18, and an unnamed feature south of the main one that surged to ~1 TW, placing all of them among the top 10 most powerful hot spots observed on Io. A temperature analysis of the features supports a simultaneous onset of these brightenings and suggests a single eruptive event propagating beneath the surface across hundreds of kilometers, the first time this has been observed on Io. This implies a connection among the hotspots' magma reservoirs, while other nearby hotspots that have been known to be active in the recent past, such as Kurdalagon Patera, appear unaffected. The simultaneity supports models of massive, interconnected magma reservoirs. The global scale of this event involving multiple hotspots and covering several hundred thousand km² should be considered in the future models of the lithosphere and interior of Io.

# Introduction

Much of the surface of Io, the most volcanic world in our solar system, is dominated by vast lava flows and paterae. Paterae are volcanic depressions resembling terrestrial calderas but often significantly larger (Carr et al., 1998; McEwen et al., 2000); however, their formation mechanism is unlikely to be the same as on Earth (Radebaugh et al., 2001). Different models have been proposed, such as (Keszthelyi et al., 2004; Radebaugh et al. 2004), where Io's paterae are surface expressions of subsurface magma intrusion processes.

Another unresolved question is the extent to which Io's heat is transported to the surface through conduction versus advection through high-temperature hotspots (McEwen et al., 2023). Io operates in a heat-pipe volcanic cycle, in which surface and subsurface eruptions act to drive the moon's lithosphere inward, suppressing outward heat flow from depth except through magmatic conduits (e.g., Spencer et al., 2020). Heat-pipe volcanism puts the deeper portion of the lithosphere into compression, plausibly explaining Io's global distribution of tall, tectonic mountains (O'Reilly and Davies, 1981). This is in conflict with the persistence of Io's continual volcanic eruptions, as compression acts to close off volcanic conduits and limit eruptions. Thrust faulting and mountain formation, however, relieves these compressional stresses and can actually drive the surface lithosphere into extension, both of which can promote volcanic eruptions (Bland and McKinnon, 2016). Juno Stellar Reference Unit (SRU) images (Becker et al. 2025) are suggestive of recent volcanic activity influenced by tectonics.

The relative contributions of surface versus subsurface eruptions (i.e., pluton formation or lateral incursion into the crust through sills) is unknown, and the interplay between Io's prodigious volcanic output and its tectonic (i.e., non-volcanic) mountain building is poorly understood (Schenk et al. 2001; Turtle et al. 2001; Jaeger et al. 2003). Understanding how much conducted heat originates from cooling surface or near-surface lava flows versus that from exposed lava, such as in lava lakes (Lopes et al. 2004, Mura et al. 2025), is crucial for modeling Io's global heat flow and regional thermal patterns and understanding its volcano-tectonic system. Observational evidence, such as the presence of numerous caldera-like depressions and long-duration eruptions like those at Prometheus, points to the existence of subsurface reservoirs or magma chambers (Leone et al., 2009; Hamilton et al. 2013).

Io offers critical insights into tidal heating, one of the primary drivers of geological activity in our solar system and beyond. Recent measurements of Io's tidal response (Park et al. 2024) indicate that while the existence of a deep magma ocean cannot be ruled out, this could not be the source of Io's volcanism. Instead, localized subsurface reservoirs, the level of connectivity of which is not well understood, may better describe the source of Io's volcanic activity. Investigating the characteristics of Io's magma chambers and transport systems, as well as monitoring Io's volcanic activity and determining recharge times and eruption intervals, is essential to advance our understanding of this enigmatic moon.

## Observations and results

On December 27, 2024, between 00:43 UTC and 01:56 UTC, the JIRAM infrared experiment on board Juno (Adriani et al., 2017; Mura et al., 2018) observed the southern hemisphere of Io, initially from a distance of 100,000 km, decreasing to 75,000 km, over about one hour, achieving a spatial resolution from 23 to 17 km. The instrument, equipped with a dual-band imager (L band: ~3.3 to 3.6 μm; M band: ~4.5 to 5 μm) and a spectrometer channel (2 to 5 μm), detected an event of extreme infrared radiance over a vast region centered at 73°S, 140°E, south and east of Nemea Planum, Crimea Mons and Olafat Patera, west of Illyrikon Regio (Figure 1, feature "A"). Such intense radiance observed by JIRAM comes from a region that covers an area of approximately 65,000 km² on Io's surface, spanning more than 200 km in latitude and more than 400 km in longitude. The radiance is so strong that it produced considerable saturation on the detector. Detached regions of similarly intense radiation, likely associated with the same event (based on thermal analysis as described in the "Methods" section), were also detected, with very high radiance as well. A very large feature ("B" in Figure 1) is just south of the most intense feature ("A" in Figure 1); at least a couple of hot spots were also identified north of the primary spot, which were previously detected (Veeder et al., 2015, see "Methods") but increased their power output by more than a factor 100. As seen in Figure 1, panel D, feature "A" could indeed consist of two distinct features, one to the east and one to the west. The signal maintained its extreme intensity throughout the observation period, suggesting the event likely persisted both before and after the observation window. This region lacked substantial thermal emission just two months earlier (orbit 66, Figure 1 panel B.). Further, images do not reveal a clear preexisting patera at this location ("A"), only previous long lava flows, though the image resolution in that region is poor.

Analyzing fully saturated images is challenging; however, the spectrometer has been used to retrieve the total power of the feature; also, the signal's strength caused a distinctive stray-light feature, attributable to the same source, to be detected in a different region of the detector (Figure 2), which is not saturated and it is used to retrieve the radiance of the saturated data (see "Methods").

Based on the analysis performed with both the spectrometer and the imager, the total power from the main feature ranges from 140 to 260 TW, making this event likely the most intense ever recorded on Io, likely exceeding the brightest event observed at Surt in 2001 (~80 TW) (Marchis et al., 2002).

What makes this event truly remarkable is the vast extent of the surface area involved. As noted, the main feature spans approximately 65,000 km² (see "Methods"), being three times larger than Loki Patera (~20,000 km$^2$), known as the largest active volcanic patera on Io (typically emitting ~10–20 TW; Rathbun et al., 2004; de Kleer and de Pater, 2017). Within the observed region, there is one feature previously identified as a dark patera by Veeder et al. (2012) and designated Pfd454. Given the sheer size of the main spot observed now, it cannot be conclusively attributed to Pfd454. However, this feature, located at 74.7°S, 134.4°E, was earlier estimated to cover an area of 300 km² with a total power output of 34 GW (Veeder et al., 2015; see "Methods"). Notably, the total power detected by JIRAM is more than 1,000 times greater than the estimated value of Pfd454 before the outburst began, highlighting the remarkable intensity of this event at a geological site previously not expected to host such a bright eruption.

Another highly intriguing factor, which could improve our understanding of volcanism on Io, is the dramatic increase in power output observed at several hot spots near the main feature ("A"). Table 1 provides a summary of the total power for different features discussed here, together with previous estimations (see "Methods" for details on the calculation). Specifically, the unnamed hot spot to the south ("B"), the Illyrikon A spot, and two hot spots to the north (PV18 and P139) have exhibited remarkable power output increases. Feature B was not previously reported and was completely dark in PJ66. In PJ66 Illyrikon A was measured to have a total power of 1.5 GW and emitted 50 GW during the event in this study. P139 and PV18 had previously measured power outputs of <80 GW and ~90 GW, respectively (the latter value is from Veeder et al., 2015), but in the JIRAM PJ68 observations both had enormous power outputs of more than 1 TW (see "Methods"). Given that these hot spots are located approximately 200 km from the main hot spot in different directions and considering that previous surveys reported no such drastic power output variations (Mura et al., 2024), it is reasonable to assume that these events, and hence the hotspots, are geologically coupled or connected. On the other hand, several other distal hotspots, such as Kurdalagon Patera, exhibit the same radiance as one month before (PJ66, see Figure 1 - panel B), or no radiance at all (P137) suggesting that this phenomenon is confined within a finite perimeter. A temperature analysis of Feature A, Feature B, PV18, and P139 (see Table 1) suggests that the temperature is similar (between 500 and 600 K); this could be explained by onset of the different brightenings happening at the same time, about one day before the observation if we apply the model by de Kleer and de Pater (2017, their figure 1). Illyrikon A shows a higher temperature, suggestive of an even more recent outburst of lava.

**Discussion**

As an explanation of the simultaneous brightening of multiple hot spots across such a wide region, we propose that the eruptions may be originating from a magma chamber or reservoir that feeds individual eruption sites via a network of sills and dikes, analogous to – but at a larger scale than – that inferred to exist beneath Hawaii (Wilding et al. 2023). It is plausible that some magma reservoirs may be so large that they can feed paterae that are significantly distant from each other, and still the regions of the interconnected patera have well-defined boundaries. We suggest that one interconnected reservoir underlies multiple hotspots and that during this time, a single large event caused magma in that reservoir to spill simultaneously over into these interconnected hotspots, possibly triggered by pressure changes in the deeper feeder reservoir, dramatic infilling from deeper within Io, or by substantial tectonic activity causing a roof collapse. This model does not require access to a global magma ocean, but can instead be explained by an extensive regional magma reservoir of significant size. Thus, the model is consistent with the interior model based on Juno gravity data proposed by Park et al. (2024), which indicates that Io does not have a global magma ocean.

The observations by Mura et al. (2024b, 2025) show that many hot spots on the surface of Io have ring-shaped thermal signatures, suggesting that these are, in truth, lava lakes on the surface. The central parts of these lakes are covered by crust, and only the lava close to the patera's walls is exposed—due to abrasion—and emitting heat. In the case of a violent event on such a lake, we would expect the crust to break apart, followed by the slow cooling of new crust and the formation of a brighter new 'ring-shaped' thermal signature. This is in fact what we observe for P139 (Figure 3, right panel) if we stretch the color scale with respect to Figure 1: the ring-shaped emission is visible, but it is much less pronounced compared to the many other lava lakes described by Mura et al. (2025), suggesting that the feature is newly formed. Stretching the color scale for Feature B (in Supplementary figure S5) shows something similar: it appears to be a lava lake with an outflow going to the west. This implies that the event triggered some changes in the equilibrium of pre-existing lava lakes (such as P139) or created new ones (Feature B), causing lava to spill out. Figure 4 shows a sketch of the proposed model.

Evidence for other past overflow events can be seen in cooled lava flows surrounding many other paterae across Io, including Loki.

On Earth, deep-seated pulses of magma can lead to subsequent volcanic eruptions, similar to the "surge" documented beneath Hawaii (Burgess and Roman, 2021), which was followed by the eruptions of two volcanoes in 2022. A similar process, on different time scale, may be at play at Io, where a deeper magma reservoir could have supplied material to several shallow chambers or reservoirs, resulting in simultaneous or near-simultaneous eruptions. A sketch of this process is provided in Figure 4. The size of the implied deep magma reservoir could significantly exceed that of its terrestrial counterparts, likely due to the higher heat flow, lower gravity and rigid lithospheric lid of Io. Eruption rates inferred from the thermal power output of ~200 TW correspond to approximately 0.2 km³/hour of material erupting and cooling through 1000 K. Assuming the eruption lasted many days, the total erupted volume could be on the order of 100 km³. Such a large volume is still insufficient to induce substantial subsidence and tectonic deformation of a hypothetical 100 by 100 km wide magma chamber (the size of the main feature we observed). These eruption rates are substantially higher than typical Hawaiian eruption rates (~0.1 km³/year), implying that the magma reservoir had insufficient time to viscoelastically relax and accommodate the added volume (Jellinek & DePaolo 2003), resulting in rapid magma expulsion. To estimate the necessary dimensions of a dike capable of sustaining a magma flow rate of 0.2 km³/hr, we consider a 100 km-long dike, which corresponds to a flow rate of approximately 0.5 m²/s per unit length. Assuming a lava viscosity of 100 Pa·s, a dike width of about 2 meters would be sufficient to accommodate this flux. This relatively small width suggests that magma supply rates are unlikely to be a limiting factor. Sills are also invoked by the model by Keszthelyi et al. (2004), where mafic magma rising through Io's crust accumulates beneath a near-surface layer of frozen volatiles, forming a horizontal magma sheet (i.e. a sill), which eventually leads to the collapse of the roof and the formation of a new patera.

## Conclusion

We have observed the eruption of a large volcanic outburst on Io, with the most intense power output ever recorded. Several other large eruptions that occurred simultaneously nearby have revealed, for the first time, the structure and vast extent of the connections between different paterae on Io: a system of apparently interconnected magma reservoirs situated 100s of km from each other. An extensive subsurface magma system on Io would help to explain the observations of globally distributed lava lakes (Mura et al., 2025), where the lake surface is thought to move up-and-down. If there were regional connections between paterae, they may all respond together and similarly to forcing from a large magma source.

An eruption of this magnitude is likely to leave a long-lived signature. Previous large eruptions at Io have left varied features such as a large area thought to be pyroclastic deposits (Pillan patera eruption in 1997, see McEwen et al. 1998), large red sulfur rings, or small, maybe fissure fed lava flows (Tvashtar in 1999, see Turtle et al. 2004 and Thor in 2001, see Lopes et al., 2004). Surface changes resulting from the eruption, and the temporal evolution of the event, can be further studied with upcoming data from Juno.

Future observations by Juno may clarify the event's nature, which could have left pyroclastic deposits or lava flows akin to previous large eruptions, offering insights into Io's volcanic and tectonic regimes. Earth-based observations of this region may also be possible, although the observational geometry is extremely limited owing to the region's high latitude. Remaining questions about changes from this eruption and the interconnectedness of magma in Io's interior may not be answerable in detail without future exploration.

**Open Research**

The data used in this study comes from Juno and the Jovian Infrared Auroral Mapper (JIRAM) and is publicly available on the Planetary Database System in PDS4 format (Adriani et al., 2019). The SPICE bundles used to calibrate the viewing geometry is also publicly available in the Planetary Database System via the Navigation and Ancillary Information Facility (NAIF). The Voyager-Galileo IO data used for our analysis is available through Williams et al., (2011a) on the U.S. Geological Survey Scientific Investigations website (Williams et al., 2011b).

*Note for the editor. The PDS archive is constantly updated. For the reviewing purposes only, we also upload the data as supplementary material.*

## Acknowledgments

We thank Agenzia Spaziale Italiana (ASI) for the support of the JIRAM contribution to the Juno mission. This work is funded by the ASI–INAF Addendum n. 2016-23-H.3-2023 to grant 2016-23-H.0. Part of this work was performed at the Jet Propulsion Laboratory, California Institute of Technology, under contract with NASA.

## Corresponding author

Correspondence to Alessandro Mura (alessandro.mura@inaf.it)

## Competing interests

The authors declare no competing interests.

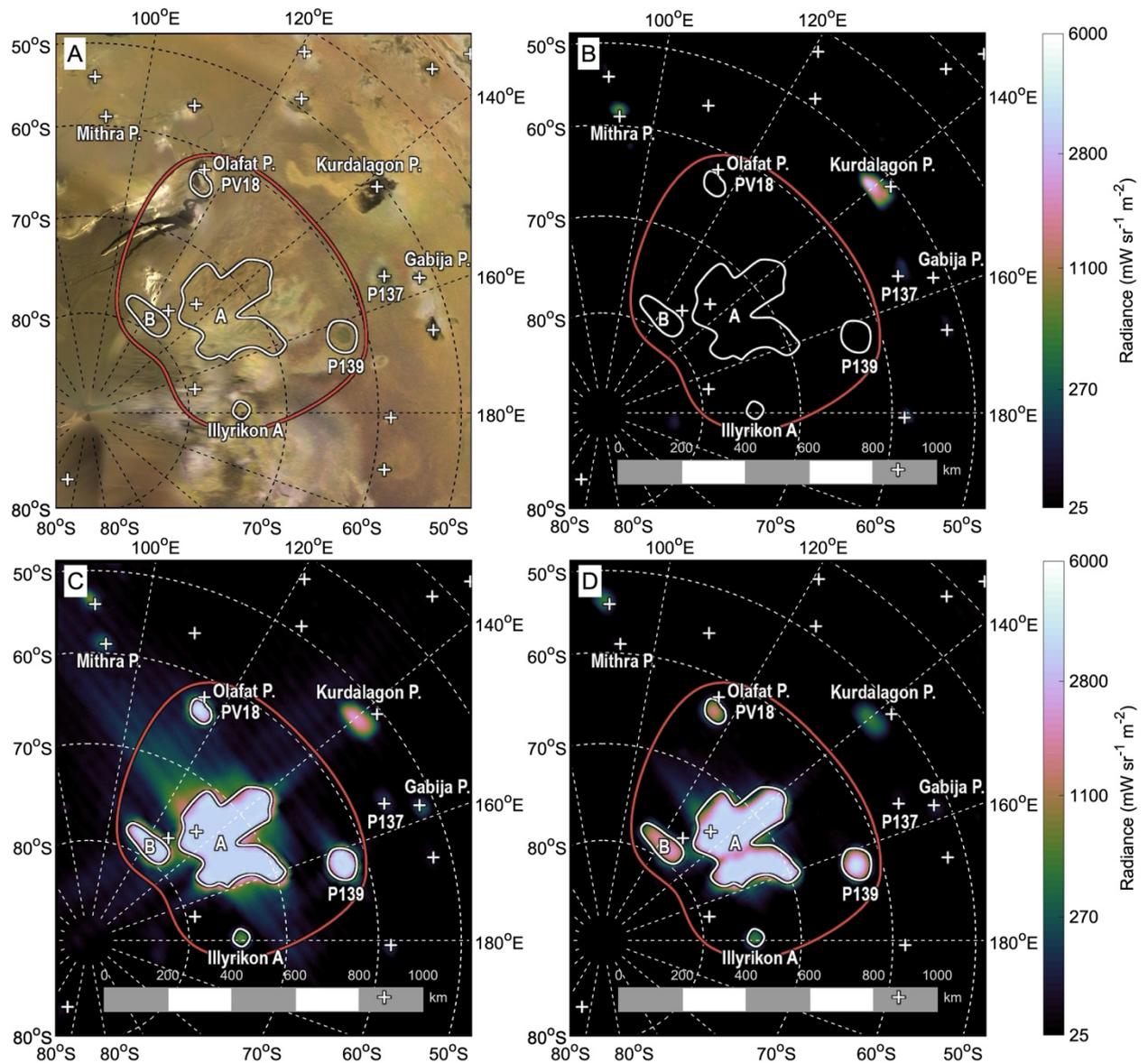

**Figure 1.** Panel A: USGS map of the region of interest (Williams et al., 2011); Panel B: radiance map in the M band, from JIRAM, taken during orbit 66 (PJ 66; Oct. 22, 2024); Panel C: same, observed during orbit 68 (PJ 68; Dec. 27, 2024); Panel D: same, for L-band data. Location of some previously known hot spots are indicated by white crosses or white contours. M-band radiance is integrated in the M band (~4.5 to 5 µm), L-band radiance is integrated in the L band (~3.3 to 3.6 µm). The color scale is not linear to enhance the retails at low radiance; white is saturated. In PJ 66, only Kurdalagon P. shows a significant radiance, and P137 is barely visible at a similar radiance as in PJ 68. PV18 P., P139, the unnamed feature at 80°S, 120°E and Illyrikon A. are completely dark in PJ 66 but visible in PJ 68. The region that encompasses the locations that erupted simultaneously, as observed in PJ 68, is drawn in red.

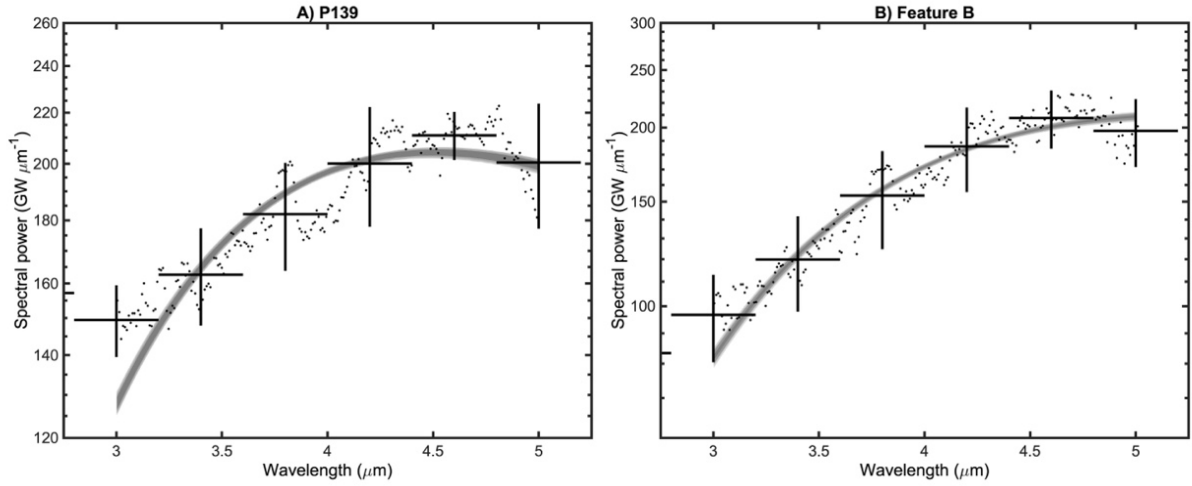

**Figure 2.** Panel A: Power spectrum from P139 and MCMC model. Dots are data, sampled at 9 nm spectral resolution; crosses are averages over 0,25 μm; gray lines are the MCMC model (increasing shades of gray show decreasing confidence levels: 95%, 90% and 75%). Panel B, same for Feature B. P139 has a model temperature of 630K ± 130 K (2-σ uncertainty), Feature B has a model temperature of 540 K ± 140 K (2-σ uncertainty).

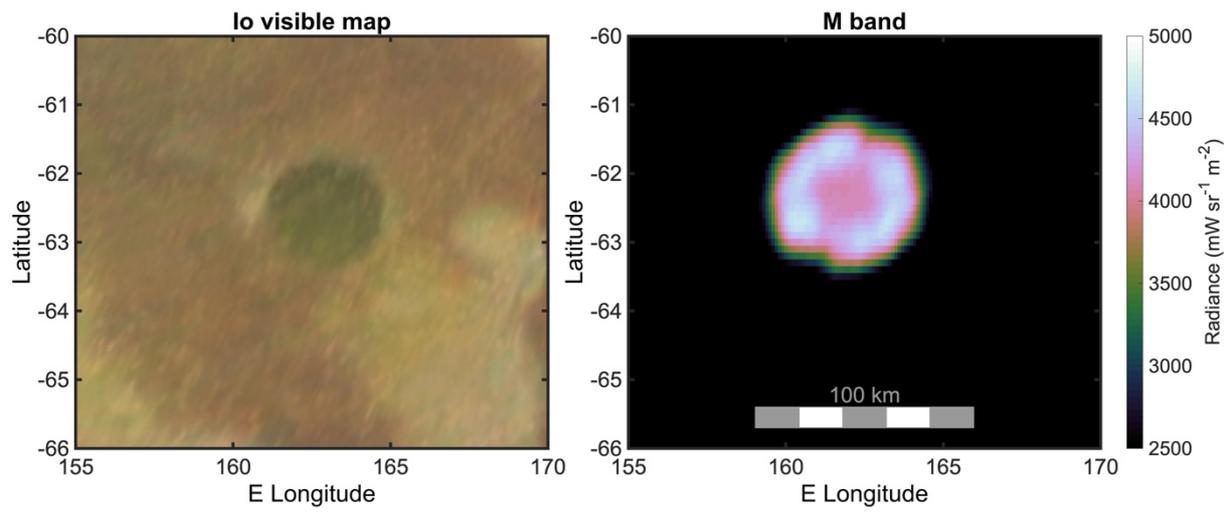

**Figure 3**. Map of P139 in the visible (left) and in the infrared (right). M-band radiance map is integrated from 4.5 to 5 µm. The IR map clearly shows a ring-shaped emission, typical of lava lakes.

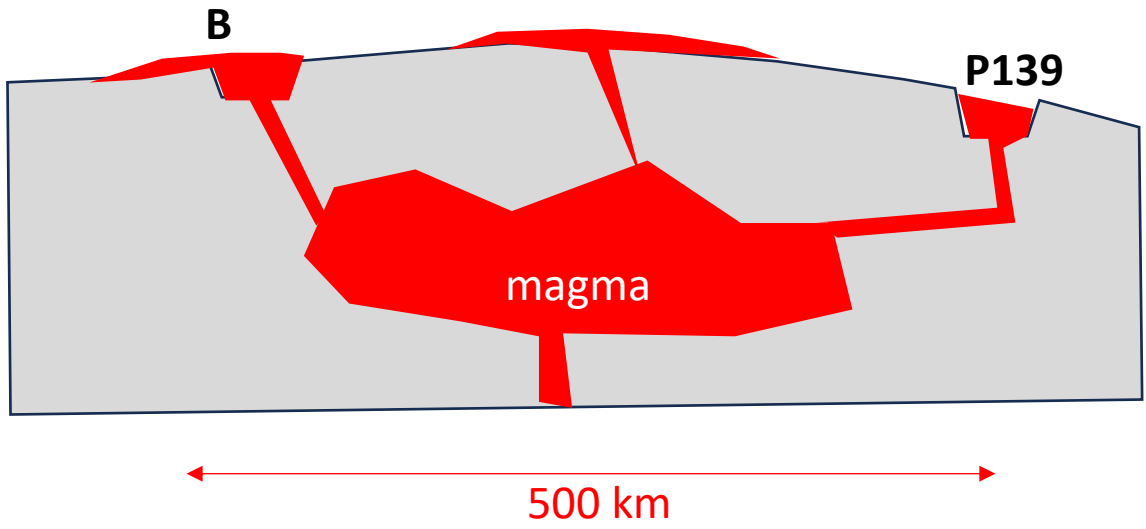

**Figure 4:** A sketch of the proposed model for the subsurface magma chamber system

| Feature | PJ66 | | | PJ68 | | | | |
|---|---|---|---|---|---|---|---|---|
| | $P_L$ | $P_M$ | $P$ | $P_L$ | $P_M$ | $P$ | $T$ | Area |
| | GW μm$^{-1}$ | GW μm$^{-1}$ | TW | GW μm$^{-1}$ | GW μm$^{-1}$ | TW | K | km$^2$ |
| A | n/a [a] | n/a [a] | 0.034 [c] | 8.9 10$^3$ | 19 10$^3$ | 200±60 | 500 ±50 | 65000 |
| B | n/a [a] | n/a [a] | | >40 | >100 | 1.5±0.2 [b] | 540 ± 150 | 300 |
| PV18 | n/a [a] | 0.35 | 0.09 [d] | 40 | 80 | ~1 | ~500 | 300 |
| P139 | n/a [a] | 0.05 | <0.08 [e] | 140 [f] | 200 [f] | 1.3±0.4 [b] | 635 ± 125 | 150 |
| Illyrikon A | 0.15 | 0.2 | 0.0015 | 10 | 7 | 0.05 | ~900 | n/a [g] |

**Table 1.** Spectral power, total power, retrieved temperature and area for 5 features discussed in this study. [a] no signal detected above the background level. [b] imager is saturated; total power is obtained with a MCMC model on the spectrometer data. [c] literature value for Pfd454, from Veeder et al. (2015). [d] from literature, Veeder et al. (2015). [e] basing on $P_M$ only, and assuming any temperature > 200 K. [f] based on the spectrometer. [g] area is too small (a few km$^2$) to perform a retrieval.